\documentclass[aps,prc,superscriptaddress,showpacs,twocolumn,floatfix]{revtex4}

\usepackage{epsfig,longtable}
\usepackage{graphicx}
\usepackage{dcolumn}
\usepackage{bm}

\def\gtorder{\mathrel{\raise.3ex\hbox{$>$}\mkern-14mu
 \lower0.6ex\hbox{$\sim$}}}
\def\ltorder{\mathrel{\raise.3ex\hbox{$<$}\mkern-14mu
 \lower0.6ex\hbox{$\sim$}}}
\def\etal{\textit{et al.}}

\begin{document}

\title{Experimental constraints on non-linearities induced by two-photon
effects in elastic and inelastic Rosenbluth separations}

\author{V. Tvaskis}
\affiliation{Hampton University, Hampton VA 23668, USA}
\affiliation{Thomas Jefferson National Accelerator Facility, Newport News, VA 23602, USA}
\author{J. Arrington}
\affiliation{Argonne National Laboratory, Argonne, IL 60439, USA}
\author{M. E. Christy}
\affiliation{Hampton University, Hampton VA 23668, USA}
\author{R. Ent}
\affiliation{Thomas Jefferson National Accelerator Facility, Newport News, VA 23602, USA}
\author{C. E. Keppel}
\affiliation{Hampton University, Hampton VA 23668, USA}
\affiliation{Thomas Jefferson National Accelerator Facility, Newport News, VA 23602, USA}
\author{Y. Liang}
\affiliation{Hampton University, Hampton VA 23668, USA}
\affiliation{American University, Washington, D.C. 20016}
\author{G. Vittorini}
\affiliation{Eckerd College, St Petersburg, FL 33711}

\date{\today}

\begin{abstract}

The effects of two-photon exchange corrections, suggested to explain the
difference between measurements of the proton elastic electromagnetic form
factors using the polarization transfer and Rosenbluth techniques, have been
studied in elastic and inelastic scattering data. Such corrections could
introduce $\varepsilon$-dependent non-linearities in inelastic Rosenbluth
separations, where $\varepsilon$ is the virtual photon polarization parameter.
It is concluded that such non-linear effects are consistent with zero for
elastic, resonance, and deep-inelastic scattering for all $Q^2$ and $W^2$
values measured.

\end{abstract}

\maketitle

\section{Introduction}\label{section:intro}

Electron-proton ($e$--$p$) scattering has proven to be a powerful tool in the
investigation of the structure of the nucleon.  This interaction is typically
described as the exchange between the electron and the proton of a single
virtual photon.  Because the electron is a point-like particle, any structure
observed in $e$--$p$ scattering must be related to the target structure.
Moreover, the relatively small value of the electromagnetic coupling constant
ensures that corrections to the one-photon exchange approximation should be
relatively small.  To further improve on the already impressive accuracy
achieved in the analysis of electron scattering data, the contribution of
two-photon exchange (TPE) effects in elastic $e$--$p$ scattering are under
theoretical investigation~\cite{maximon00, blunden03, chen04, afanasev05,
rekalo04, blunden05, kondratyuk05}. Two-photon exchange effects have garnered
particular attention as of late due to their potential role in resolving the
discrepancy between electromagnetic form factors measured through the
Rosenbluth Separation method~\cite{walker94,arrington03a,christy04,qattan05}
and a polarization transfer technique~\cite{jones00,gayou02} (see
Sect.~\ref{section:methods}).

Data from elastic and inelastic scattering, both in the resonance and
deep-inelastic regimes, are here studied in light of this concern. There is a
newly expanded, substantial data set which enables in particular a search for
non-linearities caused by TPE effects.  While dedicated measurements have been
proposed for elastic data~\cite{e05017}, and a model-dependent analysis of
non-linearity has been performed for elastic $e$--$p$
scattering~\cite{tomasi05}, this work represents a first detailed,
model-independent study of non-linear effects in elastic and inelastic
$e$--$p$ scattering data.

\section{Rosenbluth Separation Technique}\label{section:ros}

The differential cross section for $e$--$p$ scattering can be expressed in the
Born approximation in terms of absorption of longitudinal ($\sigma_L$) and
transverse ($\sigma_T$) virtual photons as
\begin{equation}
{ d^2\sigma \over d\Omega dE^{'} } = \Gamma
\Big[\sigma_T(W^2, Q^2)+\varepsilon \sigma_L(W^2, Q^2)\Big] \phantom{l},
\label{eq:roz1}
\end{equation}
where $Q^2$ is the negative squared mass of the virtual photon, $W^2$
is the mass squared of the undetected system, and $\Gamma$ is the transverse
virtual photon flux:
\begin{equation}
\Gamma =   { \alpha K \over 2\pi^2 Q^2 } { E'\over E}
{1 \over 1-\varepsilon} \phantom{l}.
\label{eq:gamma}
\end{equation}
Here, $\alpha$ is the fine structure constant, $E$ and $E'$ are the energy of
the initial and scattered electron, respectively, and $K$ is
\begin{equation}
K = {2M\nu - Q^2 \over 2M} \phantom{l},
\end{equation}
where $M$ is the mass of the proton and $\nu = E - E'$. The variable
$\varepsilon$ is the relative longitudinal virtual photon flux. Therefore
$\varepsilon$ = 0 corresponds to a purely transverse photon polarization.

The Rosenbluth separation technique is used to separate the longitudinal and
transverse components of the cross section.  Here, Eq. (\ref{eq:roz1}) is
written in the following form:
\begin{equation}
{1 \over \Gamma }{d^2\sigma \over d\Omega dE'} =
\sigma_T(W^2, Q^2)+\varepsilon\sigma_L(W^2, Q^2) \phantom{l}.
\label{eq:rozenb} 
\end{equation}
In the Born approximation, the left hand side, the \textit{reduced cross
section}, depends linearly on $\varepsilon$.  To perform the Rosenbluth
separation, data covering a range in $\varepsilon$ at fixed $(W^2, Q^2)$
values must be obtained. Any deviation from linearity must come from higher
order terms that are not included in the standard radiative correction
procedures.

\section{Two Methods of Form Factors Measurements and two Different Results}\label{section:methods}

For the case of elastic scattering, the Rosenbluth separation technique is
used to extract the form factors $G_E$ and $G_M$, from the $\varepsilon$
dependence of a reduced elastic cross section $\sigma_r$ at fixed $Q^2$, i.e.
\begin{equation}
\sigma_r \equiv \frac{d\sigma}{d\Omega}
\frac {\varepsilon (1 + \tau)}{\sigma_{Mott} } =
{\tau {G_M^2}(Q^2) + \varepsilon {G_E^2}(Q^2)} \phantom{l},
\label{eq:cs5} 
\end{equation}
where $\tau = Q^2/4M^2$.

At fixed $Q^2$, the form factors $G_E$ and $G_M$ can be extracted from a
linear fit in $\varepsilon$ to the measured reduced cross sections. Such a
Rosenbluth fit yields $\tau G_M^2$ as the intercept and $G_E^2$ as the slope.
With increasing $Q^2$, the cross section is dominated by $\tau G_M^2$, while
the relative contribution of the $G_E^2$ term is diminished. Precise
understanding of the $\varepsilon$-dependence in the radiative corrections
becomes crucial at high values of $Q^2$. Therefore, in order to measure the
ratio $G_E/G_M$ at high values of $Q^2$, a polarization transfer method
has also been employed in Hall A at Jefferson Lab (JLab).

In polarized elastic electron-proton scattering, the longitudinal and
transverse components of the recoil polarization are sensitive to different
combinations of the electric and magnetic elastic form
factors~\cite{akheizer74, arnold81}. The ratio of the form factors can be
directly related to the components of the recoil polarization
\begin{equation}
{ G_E \over G_M} = { P_t \over P_l}
{ (E+E') \tan (\theta_e / 2) \over 2M} \phantom{l},
\label{eq:pol} 
\end{equation}
where $P_l$ and $P_t$ are the longitudinal and transverse components of
the final proton polarization, and $\theta_e$ is the angle between the
initial and final directions of the lepton.

Recent measurements from Jefferson Lab using the polarization transfer
technique to measure the ratio $G_E/G_M$ have found that $G_E$ decreases
more rapidly than $G_M$ at large $Q^2$~\cite{jones00,gayou02}. This
differs from results obtained at SLAC in a similar $Q^2$ range using the
Rosenbluth technique. There exist but two explanations for this discrepancy.
There is either an unidentified systematic experimental uncertainty in the
polarization transfer data, or a systematic uncertainty common to all
Rosenbluth data.

It has been estimated that a 5-7\% systematic correction to the $\varepsilon$
dependence of the reduced Rosenbluth cross section measurements would be
needed in order to resolve the discrepancy~\cite{arrington04a, guichon03,
arrington04d}. However, a detailed analysis does not show any inconsistencies
in the cross section data sets~\cite{arrington03a}. Moreover, new high $Q^2$ cross
section data from Jefferson Lab~\cite{christy04, qattan05} are
consistent with the older SLAC cross section data~\cite{andivahis94} obtained
in the same $Q^2$ range. The results of Ref.~\cite{qattan05}, where the struck
proton rather than the scattered electron was detected, have a precision
comparable to the polarization transfer measurements. It must be concluded,
then, that the needed 5-7\% $\varepsilon$-dependent correction is not due to
standard experimental considerations in the measured Rosenbluth cross sections.

It has been suggested~\cite{guichon03, blunden03} that the
discrepancy may be explained by TPE effects not fully accounted
for in the standard radiative corrections procedure of Mo and
Tsai~\cite{mo69}.  The polarization transfer technique involves a ratio of
cross sections, and hence is expected not to be very sensitive to such
effects~\cite{maximon00, afanasev05, blunden05}. In contrast, these
contributions can significantly affect the Rosenbluth separation technique.
TPE contributions can be independent of $\varepsilon$ (affecting both $G_E^2$
and $G_M^2$ in Eq. (\ref{eq:cs5})), linear in $\varepsilon$ (significantly
affecting $G_E^2$), or non-linear in $\varepsilon$.

The experimental evidence for significant TPE contributions to the form factor
measurements is still quite limited.  While the non-zero transverse beam spin
asymmetry~\cite{wells01, maas05} provides direct evidence for TPE in elastic
$e$--$p$ scattering, we are lacking similar evidence for such effects on the
unpolarized cross sections.  The discrepancy between polarization transfer and
Rosenbluth extractions of $G_E/G_M$ provides only an indirect indication of a
missing correction, while direct searches for TPE through the comparison of
electron--proton and positron--proton scattering yield some evidence of
deviations from the Born approximation at low $\varepsilon$, but only at the
three sigma level~\cite{arrington04b}. Observation of a deviation from
linearity in the reduced cross section would provide a clear indication of TPE
(or other higher order corrections not included in standard radiative
correction procedures), though only the non-linear portion of the correction
could be directly isolated. New high-precision Rosenbluth data in
elastic~\cite{qattan05} and inelastic~\cite{liang05} $e$--$p$ scattering allow
for a much more sensitive search for such non-linearities.

This work reports results of a search for effects of TPE corrections in
elastic and inelastic scattering data by searching for $\varepsilon$-dependent
non-linearities in existing Rosenbluth separations. We note that this analysis
will not be sensitive to either systematic shifts in the reduced cross section
of Eq. (\ref{eq:rozenb}), or to two-photon effects which are linear in
$\varepsilon$.

\section{Data Overview}

Table~\ref{tab:experiments} lists the data sets included in the present
analysis.  We include several measurements of elastic $e$--$p$ to cover a
range in $Q^2$, while the SLAC measurements~\cite{dasu88b,dasu94}
and the recent JLab measurement~\cite{liang05} cover the DIS and resonance
regions.

\begin{table}
\caption{Summary of experiments included in the analysis, including the
number of L--T separations and the typical cross section uncertainties
(excluding normalization uncertainties).
\label{tab:experiments}}
\begin{tabular}{|l|c|c|c|c|}
\hline
				& $Q^2$         & \# of	& Typ.	& Lab	\\
Elastic data			& [(GeV/c)$^2$] & L--Ts	&~$\delta \sigma /\sigma$~	& \\ \hline
Janssens \etal~\cite{janssens66}  	& 0.2--0.9	& 20	& 4.7\% & Mark III	\\
Litt \etal~\cite{litt70}         	& 2.5--3.8	& 4	& 1.7\%	& SLAC	\\
Berger \etal~\cite{berger71}      	& 0.4--1.8	& 8	& 2.6\%	& Bonn	\\
Walker \etal~\cite{walker94}~$^a$  	& 1.0--3.0	& 4	& 1.1\%	& SLAC	\\
Andivahis \etal~\cite{andivahis94}~$^b$ & 1.8--5.0	& 5	& 1.3\%	& SLAC	\\
Christy \etal~\cite{christy04}		& 0.9--5.2	& 7	& 1.3\%	& JLab	\\
Qattan \etal~\cite{qattan05}~$^c$	& 2.64--4.1	& 3	& 0.6\%	& JLab	\\ \hline
 & & & & \\
Inelastic data			& $W^2$[GeV$^2$]     & 	& 	& \\ \hline
Liang \etal~\cite{liang05}		& 1.3--3.9	& 191	& 1.7\% & JLab \\
Dasu \etal~\cite{dasu88b,dasu94}	& 3.2--30 	& 61	& 3.0\% & SLAC \\ \hline
\multicolumn{5}{|l|}{$^a$ Data below $20\deg$ are excluded.} \\
\multicolumn{5}{|l|}{$^b$ Data from 8 GeV spectrometer.} \\
\multicolumn{5}{|l|}{$^c$ Excludes ``slope'' systematic uncertainties.} \\ \hline
\end{tabular}
\end{table}

For elastic $e$--$p$ scattering, we examine Rosenbluth extractions from several
different experiments.  We study the Rosenbluth separation for the experiments
and $Q^2$ values listed in Table II of Ref.~\cite{arrington03a}, including
the updated radiative corrections~\cite{arrington03a}.  In addition,
data from two recent Jefferson Lab measurements~\cite{christy04, qattan05} are
included. In all cases, the reduced cross sections are taken from a single
experiment and single detector.  Where necessary, cross section values at
slightly different $Q^2$ values are shifted to a fixed $Q^2$ values. Only small
corrections were needed, typically below 2\%, although a handful of points were
corrected by 5--10\%.  There are a total of 51 Rosenbluth separations
that we will examine for non-linearities.  Typical point-to-point
uncertainties are roughly 1--2\% for most of the data sets, although several
of the older experiment had larger uncertainties and the E01-001
results~\cite{qattan05} have point-to-point uncertainties below one percent.

\begin{figure}
\includegraphics[width=8.0cm]{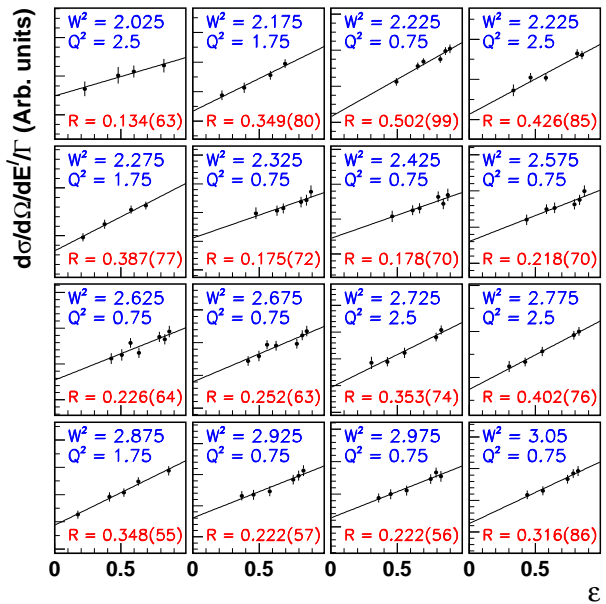}
\caption{ (Color online) Example Rosenbluth separations performed in
experiment E94-110. Each figure includes the kinematics, $W^2$ in GeV$^2$ and
$Q^2$ in (GeV/c)$^2$, and the extracted value of $R=\sigma_L/\sigma_T$.}
\label{fig:roz_sep}
\end{figure}

For the resonance region, we used newly obtained data from JLab Hall C
experiment E94-110~\cite{liang05, liangphd}, which was utilized to separate
the longitudinal and transverse unpolarized proton structure functions in the
nucleon resonance region via the Rosenbluth separation technique. The
experiment ran with seven different energies ranging from 1.2~GeV to
5.5~GeV over a scattering angle range
12.9$<$$\theta_e$$<$79.9. The total point-to-point uncertainty on the
cross section measurements was approximately 2\%~\cite{liangphd}. The data
taken from this experiment were used to perform 191 Rosenbluth separations
covering the kinematic region 0.5$<$$Q^2$$<$5.0~(GeV/c)$^2$ and 1.1$<$$W^2$$<$4.0~GeV$^2$.
Examples of these Rosenbluth separations are shown in
Fig.~\ref{fig:roz_sep}. These data were used to extract the ratio, $R$, of
longitudinal to transverse cross section components.  Rosenbluth separations
are performed in five $Q^2$ bins and 43 $W^2$ bins.  The cross section values
are interpolated to the central $W^2$ and $Q^2$ values of each bin using a
global fit to the world's resonance region data~\cite{liang05, niculescu00b},
with constraints built in to provide a smooth transition to the DIS region
and the $Q^2 \to 0$ limit.
Typical corrections were 5\%, and data that required corrections
larger than 60\% were excluded.  While a few points had large enough
corrections that the uncertainty in the interpolation may not be negligible
for the given data set, their effect on the extracted values of $P_2$ should
be small and random, thus providing a negligible contribution to the
uncertainty in the extracted limits on $P_2$.

\begin{figure}
\includegraphics[width=8.0cm]{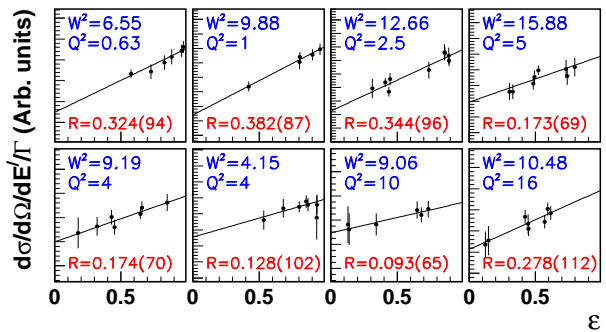}
\caption{(Color online) Example Rosenbluth separations performed
in experiment E140 at SLAC~\cite{dasu88b, dasu94}. Each figure includes
the kinematics, $W^2$ in GeV$^2$ and $Q^2$ in (GeV/c)$^2$, and the extracted
value of $R=\sigma_L/\sigma_T$.}
\label{fig:roz_sep2}
\end{figure}

For the deep-inelastic scattering (DIS) region, the data from experiment E140
at SLAC~\cite{dasu88b, dasu94} have been used. A total of 61 Rosenbluth
separations have been performed covering the kinematic region
0.63$<$$Q^2$$<$20~(GeV/c)$^2$ and 2.5$<$$W^2$$<$30~GeV$^2$. The total
point-to-point uncertainties on the cross section measurements depends on the
actual kinematics, but are typically 2--3\%. Example Rosenbluth separations
from E140 are shown in Fig.~\ref{fig:roz_sep2}.

The data from the JLab E94-110 and SLAC E140 experiments are by far the
largest data sets available for inelastic Rosenbluth separations, while the
new elastic measurements from JLab E01-001 provide significantly more precise
data for elastic scattering.

\section{Analysis and Results}

From the discussions of Sections~\ref{section:ros} and~\ref{section:methods}
it is clear that the linearity of the reduced cross section in Eq.
(\ref{eq:rozenb}) is a crucial component of the Rosenbluth technique, and that
two-photon exchange corrections could introduce a non-linear
$\varepsilon$-dependence in Eq. (\ref{eq:rozenb}). Therefore, such corrections
could manifest themselves as non-linearities in Figs.~\ref{fig:roz_sep}
and~\ref{fig:roz_sep2}.

To search for such non-linearities, the following analysis has been performed.
For each data set with three or more $\varepsilon$ values at fixed $Q^2$ and
$W^2$, the reduced cross sections are fit to a quadratic in $\varepsilon$,
of the form
\begin{equation}
\sigma_r = P_0 \cdot [ 1 + P_1 (\varepsilon-0.5) + P_2 (\varepsilon-0.5)^2 ].
\label{eq:quadfit}
\end{equation}
In the absence of TPE, we expect $P_0 = \sigma_T + 0.5 \sigma_L$, $P_1 =
\sigma_L$, and $P_2 = 0$.  TPE corrections can modify $P_0$ and $P_1$, and may
introduce a non-zero value of $P_2$, the fractional curvature relative to the
$P_0$, the cross section at $\varepsilon=0.5$. The only estimates we have for
the size of the non-linearity come from calculations for elastic $e$--$p$
scattering.  If one takes the calculations~\cite{blunden03, chen04} of TPE
effects for elastic scattering and scales the size of the corrections so that
they are large enough to explain the discrepancy between polarization and
Rosenbluth extractions, as done in~\cite{e05017}, one obtains $P_2$ values of
$\approx$6--9\%, although the precise value depends significantly on $Q^2$ and
the $\varepsilon$ range of the data.

While $P_2$ represents the fractional curvature, the size of cross section
deviations from linearity will be much smaller. For $P_2 = 10$\%, the maximum
deviation of the cross section from $P_2=0$ would be 2.5\%, at $\varepsilon =
0,1$. The effects are even smaller if the $\varepsilon$ range of the data,
$\Delta \varepsilon$, is less than one.  In this case, the deviations from
$P_2=0$ will go approximately as $(\Delta \varepsilon)^2$.  Finally, when one
performs the Rosenbluth separation, the extracted values of $\sigma_L$ and
$\sigma_T$ will be shifted from their true values in order to minimize the
deviation from the straight line fit, reducing the deviations by roughly a
factor of two from the size of the $P_2$ contribution in Eq.~\ref{eq:quadfit}.
Thus, the maximum observed deviations from linearity will be,
\begin{equation}
\Delta_{max} = \frac{(\sigma - \sigma_{fit})_{max}}{\sigma} \approx P_2 \cdot (\Delta \varepsilon)^2 / 8,
\label{eq:maxdeviation}
\end{equation}
typically more than a factor of ten smaller than the value of $P_2$.  For the
expected $P_2$ values of $\ltorder$10\% and a rather large $\Delta
\varepsilon$ range of 0.8, one expects maximum deviations from linearity to be
at the level of $\ltorder$0.8\%.  So even with high precision measurements and
a good $\varepsilon$ range, one needs a large data set to provide meaningful
limits on the non-linearities.

\begin{figure}
\includegraphics[width=8.0cm,height=5.5cm]{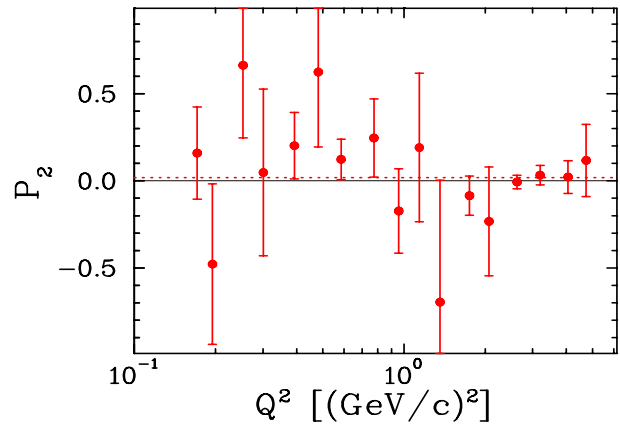}
\caption{(Color online) Extracted values of the curvature parameter, $P_2$, as
extracted from the elastic data as a function of $Q^2$.  The red dotted line
indicates the average value, $\langle P_2 \rangle = 0.019 \pm 0.027$.}
\label{fig:elastic}
\end{figure}

\begin{figure}
\includegraphics[width=8.0cm,height=5.5cm]{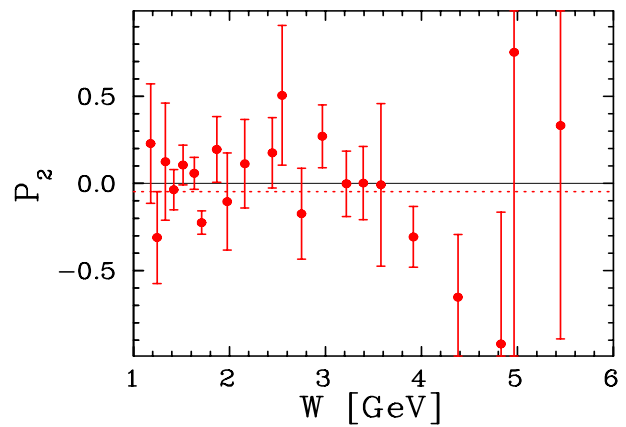}
\caption{(Color online) Extracted values of the curvature parameter, $P_2$, as
extracted from the inelastic data as a function of $W$.  Data in each $W$
bin is averaged over all $Q^2$ values in the resonance region and DIS
measurements.The red dotted line indicates the average value, $\langle P_2
\rangle = -0.048 \pm 0.036$.}
\label{fig:inelastic}
\end{figure}

We perform the fit from Eq.~\ref{eq:quadfit} for each of the elastic,
resonance region, and DIS Rosenbluth data sets.  Figures~\ref{fig:elastic}
and~\ref{fig:inelastic} show $P_2$, binned in $Q^2$ for the elastic data, and
binned in $W$ for the resonance and DIS data.  The results are consistent with
no non-linearities, and there is no apparent $Q^2$ or $W^2$ dependence. 
Table~\ref{tab:p2} shows the extracted value for $P_2$, the 95\% confidence
level upper limit on $\mid P_2 \mid$, and the approximate maximum deviation
from linearity for the elastic, resonance region, and DIS ($W^2 > 4$~GeV$^2$)
data. From these results, we determine the 95\% confidence level upper limits
on $\mid P_2 \mid of 6.4$\% for the elastic data and 10.7\% for the inelastic
data.  This yields limits on the deviations of the data from the Rosenbluth
fit of roughly 0.4\% (0.7\%) for the elastic (inelastic), assuming a $\Delta
\varepsilon$ range of 0.7.

\begin{table}
\caption{Extracted values and 95\% confidence level upper limit on $P_2$.
$\Delta_{max}$ is the upper limit on deviations of the cross section from
linearity (Eq.~\ref{eq:maxdeviation}).
\label{tab:p2}}
\begin{tabular}{|l|c|c|c|}
\hline
	& $\langle P_2 \rangle$	&$\mid P_2 \mid _{max}$ & $\Delta_{max}$\\
	&			& ~95\% C.L.~		& ~95\% C.L.~	\\
\hline
Elastic & ~0.019(27)		& 0.064		&~0.8\%$\cdot (\Delta \varepsilon)^2$~	\\
Resonance&-0.060(42)		& 0.086		& 1.1\%$\cdot (\Delta \varepsilon)^2$	\\
DIS 	& -0.012(71)		& 0.146		& 1.8\%$\cdot (\Delta \varepsilon)^2$	\\
\hline
\end{tabular}
\end{table}

Note that it is also possible for a purely linear correction to introduce a
small non-linearity, since
\begin{equation}
\frac{\sigma_r}{\sigma_T} = (1 + R \varepsilon) \cdot
(1 + C_{2\gamma} \varepsilon) =
1 + ( R + C_{2\gamma})\varepsilon + R \cdot C_{2\gamma} \varepsilon^2,
\end{equation}
where $R = \sigma_L / \sigma_T$ and $(1 + C_{2\gamma} \varepsilon)$ is the TPE
correction factor.  For the elastic data at high $Q^2$ and all of the
inelastic data presented here, $R \ltorder 0.2-0.3$, while estimates TPE
predict a change in slope, $C_{2\gamma}$, of approximately 5\%.  Hence, the
nonlinear term arising from a linear TPE correction, $R \cdot C_{2\gamma}$,
will be very small, yielding $P_2 \ltorder 1$\%.  At low $Q^2$ values, the
value of $R$ for the elastic cross section becomes quite large, yielding
values of $P_2$ on the scale of $C_{2\gamma}$ for $R \gtorder 1$.  However,
$R$ is only this large for $Q^2 \ltorder 0.4$~GeV$^2$, where the TPE
corrections decrease as $Q^2 \to 0$~\cite{blunden05, guichon03, arrington04b,
arrington04d}. The effect is $\ltorder$1\% if one assumes that $C_{2\gamma}$
increases slowly as one goes up from $Q^2=0$, as it does in
calculations~\cite{blunden05} and phenomenological extractions of the
TPE corrections~\cite{guichon03, arrington04d}. If one takes a
more rapid increase with $Q^2$, $C_{2\gamma} = 0.06 \cdot
[1-\exp(Q^2/0.5\mbox{GeV}^2)]$, we obtain values of $P_2$ coming from the
linear correction of 1.5--2.5\% for $Q^2 < 1$~GeV$^2$.  Thus, the size of this
effect should be well below the sensitivity of the existing data in all cases.

To better visualize the limits on nonlinearities, we have also performed a
global comparison of the residuals between the reduced cross sections and a
\textit{linear} fit to the reduced cross sections.  For the previous fit,
data sets with a very small $\Delta \varepsilon$ range have very little
sensitivity to the curvature.  Although these data sets have large
uncertainties, they still provide meaningful $P_2$ values. When plotting the
residuals, we want to exclude such data sets because the data points
have uncertainties comparable to the other data sets, but the residuals
little sensitivity to non-linearities.  Thus, we require include only those
data sets where $\Delta \varepsilon \geq 0.4$ when studying the residuals. 
This cut reduces the number of data sets to 35 for elastic kinematics, 77 in
the resonance region, and 38 in the DIS region.

\begin{figure}
\includegraphics[width=8.0cm,height=9.5cm]{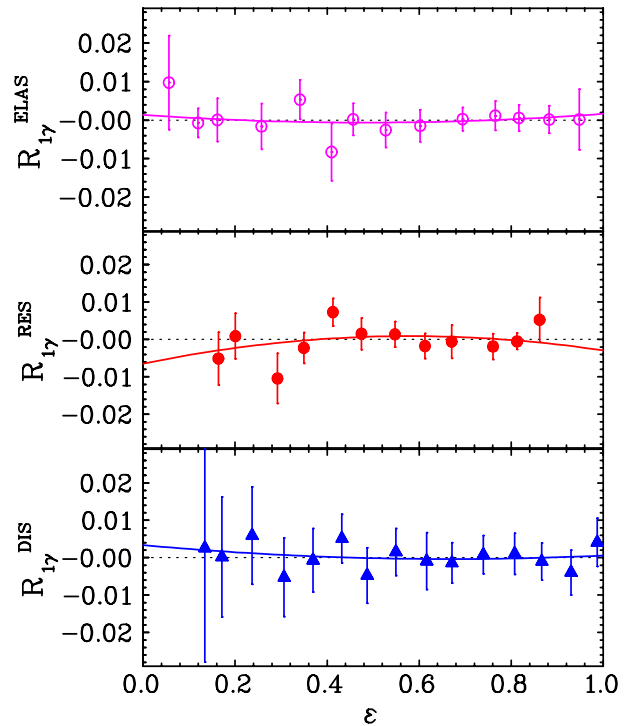}
\caption{(Color online) The weighted average of $R_{1\gamma}=(\sigma_{Data} -
\sigma_{Fit})/\sigma_{Fit}$ for the elastic measurements, the resonance region
data (JLab experiment E94-110) and the DIS data (SLAC experiment E140).
The solid lines are the fits to the form of Eq.~\ref{eq:quadfit2}.}
\label{fig:rgamma}
\end{figure}

For the data sets with sufficient $\varepsilon$ coverage, we take $R_{1\gamma}$
to be the residual from the Rosenbluth (one-photon exchange) fit,
\begin{equation}
R_{1\gamma} =  \frac{\sigma_{Data} - \sigma_{fit}}{\sigma_{fit}},
\label{eq:rgamma} 
\end{equation}
and obtain a value of $R_{1\gamma}$ for every cross section measurement in the
Rosenbluth data sets.  We can then determine the weighted average value from
the extracted $R_{1\gamma}$ values in $\varepsilon$ bins for the elastic,
resonance region, and inelastic data sets.  In the absence of TPE
contributions, one expects $R_{1\gamma} = 0$ in every $\varepsilon$ bin and,
hence, any $\varepsilon$ dependence to $R_{1\gamma}$ is an indication of
two-photon exchange. Figure~\ref{fig:rgamma} shows the combined $R_{1\gamma}$
as a function of $\varepsilon$ for the elastic, resonance region, and DIS
data, and all three data sets are consistent with $R_{1\gamma}$=0.  One can
see that while the elastic and resonance region data have high precision, the
resonance region data has less data as $\varepsilon \to 0$ and 1, while the
DIS data has limited data an low $\varepsilon$, as well as lower statistical
precision in general.

We fit the combined residuals to the form
\begin{equation}
R_{1\gamma} = A + B (\varepsilon-\varepsilon_0)^2,
\label{eq:quadfit2}
\end{equation}
where $A$, $B$, and $\varepsilon_0$ are the fit parameters.  Because we
are fitting to residuals that have already had the expected linear cross
section dependence removed, we expect that $R_{1\gamma}$ will average to zero,
yielding $A \approx 0$ in the absence of any strong $\varepsilon$ dependence.
Indeed, we find $A \ltorder 0.05$\% for the elastic, resonance, and DIS data.
The quadratic fit to $R_{1\gamma}$ yields a curvature parameter, $B$, consistent
with zero for all data sets.  We obtain $B = (0.9 \pm 2.0)$\% for the elastic
data, $(-2.3 \pm 3.0)$\% for the resonance region data, and $(0.9 \pm 3.8)$\%
for the DIS measurements.

While the limits in Tab.~\ref{tab:p2} provide the best quantitative limits
on deviations from linearity, the residuals shown in Fig.~\ref{fig:rgamma}
give a better idea of the sensitivity of the different data sets in different
regions of $\varepsilon$.  The parameterization of Eq.~\ref{eq:quadfit} assumes
a simple quadratic non-linear term, while some models for the contribution
to elastic scattering indicate larger non-linearities for $\varepsilon \to 1$.
From Fig.~\ref{fig:rgamma} we see that this region is not as well constrained
for the resonance region data, while very low $\varepsilon$ values are not well
constrained except in the elastic data.

\section{Conclusion}

We have searched for possible two-photon exchange contributions that show up
as non-linearities in Rosenbluth separations. We have used existing data in
the elastic and deep-inelastic scattering region and recent data in the
nucleon resonance region. We do not find any evidence for TPE effects. The
95\% confidence level upper limit on the curvature parameter, $P_2$, was found
to be 6.4\% (10.6\%) for the elastic (inelastic) data. This limits maximum
deviations from a linear fit to $\ltorder$0.4\% (0.7\%) for typical elastic
(inelastic) Rosenbluth separation data sets.

\begin{acknowledgments}

This work was supported in part by research grants 0099540 and 9633750
from the National Science Foundation and DOE grant W-31-109-ENG-38.
The authors wish to thank Arie Bodek for useful discussions and comments.

\end{acknowledgments}

\bibliography{2gamma}

\begin{thebibliography}{33}
\expandafter\ifx\csname natexlab\endcsname\relax\def\natexlab#1{#1}\fi
\expandafter\ifx\csname bibnamefont\endcsname\relax
  \def\bibnamefont#1{#1}\fi
\expandafter\ifx\csname bibfnamefont\endcsname\relax
  \def\bibfnamefont#1{#1}\fi
\expandafter\ifx\csname citenamefont\endcsname\relax
  \def\citenamefont#1{#1}\fi
\expandafter\ifx\csname url\endcsname\relax
  \def\url#1{\texttt{#1}}\fi
\expandafter\ifx\csname urlprefix\endcsname\relax\def\urlprefix{URL }\fi
\providecommand{\bibinfo}[2]{#2}
\providecommand{\eprint}[2][]{\url{#2}}

\bibitem[{\citenamefont{Maximon and Parke}(2000)}]{maximon00}
\bibinfo{author}{\bibfnamefont{L.~C.} \bibnamefont{Maximon}} \bibnamefont{and}
  \bibinfo{author}{\bibfnamefont{W.~C.} \bibnamefont{Parke}},
  \bibinfo{journal}{Phys. Rev.} \textbf{\bibinfo{volume}{C61}},
  \bibinfo{pages}{045502} (\bibinfo{year}{2000}).

\bibitem[{\citenamefont{Blunden et~al.}(2003)\citenamefont{Blunden,
  Melnitchouk, and Tjon}}]{blunden03}
\bibinfo{author}{\bibfnamefont{P.~G.} \bibnamefont{Blunden}},
  \bibinfo{author}{\bibfnamefont{W.}~\bibnamefont{Melnitchouk}},
  \bibnamefont{and} \bibinfo{author}{\bibfnamefont{J.~A.} \bibnamefont{Tjon}},
  \bibinfo{journal}{Phys. Rev. Lett.} \textbf{\bibinfo{volume}{91}},
  \bibinfo{pages}{142304} (\bibinfo{year}{2003}).

\bibitem[{\citenamefont{Chen et~al.}(2004)\citenamefont{Chen, Afanasev,
  Brodsky, Carlson, and Vanderhaeghen}}]{chen04}
\bibinfo{author}{\bibfnamefont{Y.~C.} \bibnamefont{Chen}},
  \bibinfo{author}{\bibfnamefont{A.}~\bibnamefont{Afanasev}},
  \bibinfo{author}{\bibfnamefont{S.~J.} \bibnamefont{Brodsky}},
  \bibinfo{author}{\bibfnamefont{C.~E.} \bibnamefont{Carlson}},
  \bibnamefont{and}
  \bibinfo{author}{\bibfnamefont{M.}~\bibnamefont{Vanderhaeghen}},
  \bibinfo{journal}{Phys. Rev. Lett.} \textbf{\bibinfo{volume}{93}},
  \bibinfo{pages}{122301} (\bibinfo{year}{2004}).

\bibitem[{\citenamefont{Afanasev et~al.}(2005)\citenamefont{Afanasev, Brodsky,
  Carlson, Chen, and Vanderhaeghen}}]{afanasev05}
\bibinfo{author}{\bibfnamefont{A.~V.} \bibnamefont{Afanasev}},
  \bibinfo{author}{\bibfnamefont{S.~J.} \bibnamefont{Brodsky}},
  \bibinfo{author}{\bibfnamefont{C.~E.} \bibnamefont{Carlson}},
  \bibinfo{author}{\bibfnamefont{Y.-C.} \bibnamefont{Chen}}, \bibnamefont{and}
  \bibinfo{author}{\bibfnamefont{M.}~\bibnamefont{Vanderhaeghen}},
  \bibinfo{journal}{Phys. Rev.} \textbf{\bibinfo{volume}{D72}},
  \bibinfo{pages}{013008} (\bibinfo{year}{2005}).

\bibitem[{\citenamefont{Rekalo and Tomasi-Gustafsson}(2004)}]{rekalo04}
\bibinfo{author}{\bibfnamefont{M.~P.} \bibnamefont{Rekalo}} \bibnamefont{and}
  \bibinfo{author}{\bibfnamefont{E.}~\bibnamefont{Tomasi-Gustafsson}},
  \bibinfo{journal}{Eur. Phys. J.} \textbf{\bibinfo{volume}{A22}},
  \bibinfo{pages}{331} (\bibinfo{year}{2004}).

\bibitem[{\citenamefont{Blunden et~al.}(2005)\citenamefont{Blunden,
  Melnitchouk, and Tjon}}]{blunden05}
\bibinfo{author}{\bibfnamefont{P.~G.} \bibnamefont{Blunden}},
  \bibinfo{author}{\bibfnamefont{W.}~\bibnamefont{Melnitchouk}},
  \bibnamefont{and} \bibinfo{author}{\bibfnamefont{J.~A.} \bibnamefont{Tjon}},
  \bibinfo{journal}{Phys. Rev.} \textbf{\bibinfo{volume}{C72}},
  \bibinfo{pages}{034612} (\bibinfo{year}{2005}).

\bibitem[{\citenamefont{Kondratyuk et~al.}(2005)\citenamefont{Kondratyuk,
  Blunden, Melnitchouk, and Tjon}}]{kondratyuk05}
\bibinfo{author}{\bibfnamefont{S.}~\bibnamefont{Kondratyuk}},
  \bibinfo{author}{\bibfnamefont{P.~G.} \bibnamefont{Blunden}},
  \bibinfo{author}{\bibfnamefont{W.}~\bibnamefont{Melnitchouk}},
  \bibnamefont{and} \bibinfo{author}{\bibfnamefont{J.~A.} \bibnamefont{Tjon}}
  (\bibinfo{year}{2005}), \eprint{nucl-th/0506026}.

\bibitem[{\citenamefont{Walker et~al.}(1994)}]{walker94}
\bibinfo{author}{\bibfnamefont{R.~C.} \bibnamefont{Walker}}
  \bibnamefont{et~al.}, \bibinfo{journal}{Phys. Rev. D}
  \textbf{\bibinfo{volume}{49}}, \bibinfo{pages}{5671} (\bibinfo{year}{1994}).

\bibitem[{\citenamefont{Arrington}(2003)}]{arrington03a}
\bibinfo{author}{\bibfnamefont{J.}~\bibnamefont{Arrington}},
  \bibinfo{journal}{Phys. Rev. C} \textbf{\bibinfo{volume}{68}},
  \bibinfo{pages}{034325} (\bibinfo{year}{2003}).

\bibitem[{\citenamefont{Christy et~al.}(2004)}]{christy04}
\bibinfo{author}{\bibfnamefont{M.~E.} \bibnamefont{Christy}}
  \bibnamefont{et~al.}, \bibinfo{journal}{Phys. Rev. C}
  \textbf{\bibinfo{volume}{70}}, \bibinfo{pages}{015206}
  (\bibinfo{year}{2004}).

\bibitem[{\citenamefont{Qattan et~al.}(2005)}]{qattan05}
\bibinfo{author}{\bibfnamefont{I.~A.} \bibnamefont{Qattan}}
  \bibnamefont{et~al.}, \bibinfo{journal}{Phys. Rev. Lett.}
  \textbf{\bibinfo{volume}{94}}, \bibinfo{pages}{142301}
  (\bibinfo{year}{2005}).

\bibitem[{\citenamefont{Jones et~al.}(2000)}]{jones00}
\bibinfo{author}{\bibfnamefont{M.~K.} \bibnamefont{Jones}}
  \bibnamefont{et~al.}, \bibinfo{journal}{Phys. Rev. Lett.}
  \textbf{\bibinfo{volume}{84}}, \bibinfo{pages}{1398} (\bibinfo{year}{2000}).

\bibitem[{\citenamefont{Gayou et~al.}(2002)}]{gayou02}
\bibinfo{author}{\bibfnamefont{O.}~\bibnamefont{Gayou}} \bibnamefont{et~al.},
  \bibinfo{journal}{Phys. Rev. Lett.} \textbf{\bibinfo{volume}{88}},
  \bibinfo{pages}{092301} (\bibinfo{year}{2002}).

\bibitem[{\citenamefont{Arrington et~al.}()}]{e05017}
\bibinfo{author}{\bibfnamefont{J.}~\bibnamefont{Arrington}}
  \bibnamefont{et~al.}, \bibinfo{howpublished}{Jefferson lab experiment
  E05-017}.

\bibitem[{\citenamefont{Tomasi-Gustafsson and Gakh}(2005)}]{tomasi05}
\bibinfo{author}{\bibfnamefont{E.}~\bibnamefont{Tomasi-Gustafsson}}
  \bibnamefont{and} \bibinfo{author}{\bibfnamefont{G.~I.} \bibnamefont{Gakh}},
  \bibinfo{journal}{Phys. Rev. C} \textbf{\bibinfo{volume}{72}},
  \bibinfo{pages}{015209} (\bibinfo{year}{2005}).

\bibitem[{\citenamefont{Akheizer and Rekalo}(1974)}]{akheizer74}
\bibinfo{author}{\bibfnamefont{A.~I.} \bibnamefont{Akheizer}} \bibnamefont{and}
  \bibinfo{author}{\bibfnamefont{M.~P.} \bibnamefont{Rekalo}},
  \bibinfo{journal}{Sov. J. Part. Nucl.} \textbf{\bibinfo{volume}{4}},
  \bibinfo{pages}{236} (\bibinfo{year}{1974}).

\bibitem[{\citenamefont{Arnold et~al.}(1981)\citenamefont{Arnold, Carlson, and
  Gross}}]{arnold81}
\bibinfo{author}{\bibfnamefont{R.~G.} \bibnamefont{Arnold}},
  \bibinfo{author}{\bibfnamefont{C.~E.} \bibnamefont{Carlson}},
  \bibnamefont{and} \bibinfo{author}{\bibfnamefont{F.}~\bibnamefont{Gross}},
  \bibinfo{journal}{Phys. Rev. C} \textbf{\bibinfo{volume}{23}},
  \bibinfo{pages}{363} (\bibinfo{year}{1981}).

\bibitem[{\citenamefont{Arrington}(2004{\natexlab{a}})}]{arrington04a}
\bibinfo{author}{\bibfnamefont{J.}~\bibnamefont{Arrington}},
  \bibinfo{journal}{Phys. Rev. C} \textbf{\bibinfo{volume}{69}},
  \bibinfo{pages}{022201(R)} (\bibinfo{year}{2004}{\natexlab{a}}).

\bibitem[{\citenamefont{Guichon and Vanderhaeghen}(2003)}]{guichon03}
\bibinfo{author}{\bibfnamefont{P.~A.~M.} \bibnamefont{Guichon}}
  \bibnamefont{and}
  \bibinfo{author}{\bibfnamefont{M.}~\bibnamefont{Vanderhaeghen}},
  \bibinfo{journal}{Phys. Rev. Lett.} \textbf{\bibinfo{volume}{91}},
  \bibinfo{pages}{142303} (\bibinfo{year}{2003}).

\bibitem[{\citenamefont{Arrington}(2005)}]{arrington04d}
\bibinfo{author}{\bibfnamefont{J.}~\bibnamefont{Arrington}},
  \bibinfo{journal}{Phys. Rev. C} \textbf{\bibinfo{volume}{71}},
  \bibinfo{pages}{015202} (\bibinfo{year}{2005}).

\bibitem[{\citenamefont{Andivahis et~al.}(1994)}]{andivahis94}
\bibinfo{author}{\bibfnamefont{L.}~\bibnamefont{Andivahis}}
  \bibnamefont{et~al.}, \bibinfo{journal}{Phys. Rev. D}
  \textbf{\bibinfo{volume}{50}}, \bibinfo{pages}{5491} (\bibinfo{year}{1994}).

\bibitem[{\citenamefont{Mo and Tsai}(1969)}]{mo69}
\bibinfo{author}{\bibfnamefont{L.~W.} \bibnamefont{Mo}} \bibnamefont{and}
  \bibinfo{author}{\bibfnamefont{Y.-S.} \bibnamefont{Tsai}},
  \bibinfo{journal}{Rev. Mod. Phys.} \textbf{\bibinfo{volume}{41}},
  \bibinfo{pages}{205} (\bibinfo{year}{1969}).

\bibitem[{\citenamefont{Wells et~al.}(2001)}]{wells01}
\bibinfo{author}{\bibfnamefont{S.~P.} \bibnamefont{Wells}} \bibnamefont{et~al.}
  (\bibinfo{collaboration}{SAMPLE}), \bibinfo{journal}{Phys. Rev. C}
  \textbf{\bibinfo{volume}{63}}, \bibinfo{pages}{064001}
  (\bibinfo{year}{2001}).

\bibitem[{\citenamefont{Maas et~al.}(2005)}]{maas05}
\bibinfo{author}{\bibfnamefont{F.~E.} \bibnamefont{Maas}} \bibnamefont{et~al.},
  \bibinfo{journal}{Phys. Rev. Lett.} \textbf{\bibinfo{volume}{94}},
  \bibinfo{pages}{082001} (\bibinfo{year}{2005}).

\bibitem[{\citenamefont{Arrington}(2004{\natexlab{b}})}]{arrington04b}
\bibinfo{author}{\bibfnamefont{J.}~\bibnamefont{Arrington}},
  \bibinfo{journal}{Phys. Rev. C} \textbf{\bibinfo{volume}{69}},
  \bibinfo{pages}{032201(R)} (\bibinfo{year}{2004}{\natexlab{b}}).

\bibitem[{\citenamefont{Liang et~al.}(2004)}]{liang05}
\bibinfo{author}{\bibfnamefont{Y.}~\bibnamefont{Liang}} \bibnamefont{et~al.}
  (\bibinfo{collaboration}{Jefferson Lab Hall C E94-110})
  (\bibinfo{year}{2004}), \eprint{nucl-ex/0410027}.

\bibitem[{\citenamefont{Dasu et~al.}(1988)}]{dasu88b}
\bibinfo{author}{\bibfnamefont{S.}~\bibnamefont{Dasu}} \bibnamefont{et~al.},
  \bibinfo{journal}{Phys. Rev. Lett.} \textbf{\bibinfo{volume}{61}},
  \bibinfo{pages}{1061} (\bibinfo{year}{1988}).

\bibitem[{\citenamefont{Dasu et~al.}(1994)}]{dasu94}
\bibinfo{author}{\bibfnamefont{S.}~\bibnamefont{Dasu}} \bibnamefont{et~al.},
  \bibinfo{journal}{Phys. Rev. D} \textbf{\bibinfo{volume}{49}},
  \bibinfo{pages}{5641} (\bibinfo{year}{1994}).

\bibitem[{\citenamefont{Janssens et~al.}(1966)\citenamefont{Janssens,
  Hofstadter, Huges, and Yearian}}]{janssens66}
\bibinfo{author}{\bibfnamefont{T.}~\bibnamefont{Janssens}},
  \bibinfo{author}{\bibfnamefont{R.}~\bibnamefont{Hofstadter}},
  \bibinfo{author}{\bibfnamefont{E.~B.} \bibnamefont{Huges}}, \bibnamefont{and}
  \bibinfo{author}{\bibfnamefont{M.~R.} \bibnamefont{Yearian}},
  \bibinfo{journal}{Phys. Rev.} \textbf{\bibinfo{volume}{142}},
  \bibinfo{pages}{922} (\bibinfo{year}{1966}).

\bibitem[{\citenamefont{Litt et~al.}(1970)}]{litt70}
\bibinfo{author}{\bibfnamefont{J.}~\bibnamefont{Litt}} \bibnamefont{et~al.},
  \bibinfo{journal}{Phys. Lett. B} \textbf{\bibinfo{volume}{31}},
  \bibinfo{pages}{40} (\bibinfo{year}{1970}).

\bibitem[{\citenamefont{Berger et~al.}(1971)\citenamefont{Berger, Burkert,
  Knop, Langenbeck, and Rith}}]{berger71}
\bibinfo{author}{\bibfnamefont{C.}~\bibnamefont{Berger}},
  \bibinfo{author}{\bibfnamefont{V.}~\bibnamefont{Burkert}},
  \bibinfo{author}{\bibfnamefont{G.}~\bibnamefont{Knop}},
  \bibinfo{author}{\bibfnamefont{B.}~\bibnamefont{Langenbeck}},
  \bibnamefont{and} \bibinfo{author}{\bibfnamefont{K.}~\bibnamefont{Rith}},
  \bibinfo{journal}{Phys. Lett. B} \textbf{\bibinfo{volume}{35}},
  \bibinfo{pages}{87} (\bibinfo{year}{1971}).

\bibitem[{\citenamefont{Liang}(2002)}]{liangphd}
\bibinfo{author}{\bibfnamefont{Y.}~\bibnamefont{Liang}}, Ph.D. thesis,
  \bibinfo{school}{American University} (\bibinfo{year}{2002}).

\bibitem[{\citenamefont{Niculescu et~al.}(2000)}]{niculescu00b}
\bibinfo{author}{\bibfnamefont{I.}~\bibnamefont{Niculescu}}
  \bibnamefont{et~al.}, \bibinfo{journal}{Phys. Rev. Lett.}
  \textbf{\bibinfo{volume}{85}}, \bibinfo{pages}{1186} (\bibinfo{year}{2000}).

\end{thebibliography}

\end{document}